\documentclass[sn-chicago]{sn-jnl}

\usepackage{amsmath, amsthm,amssymb}
\usepackage{graphicx}
\usepackage[english]{babel}
\usepackage{color}
\usepackage{url}
\usepackage{caption}
\usepackage{subcaption}

\jyear{2021}%

\theoremstyle{thmstyleone}%
\newtheorem{theorem}{Theorem}
\newtheorem{proposition}[theorem]{Proposition}%

\theoremstyle{thmstyletwo}%

\theoremstyle{thmstylethree}%

\DeclareMathOperator{\diag}{diag}

\DeclareMathOperator{\mean}{mean}
\DeclareMathOperator{\q}{q}

\def\blue#1{#1}

\newcommand{\tr}{^{\text{\tiny T}}}

\newcommand{\citeN}[1]{\cite{#1}}
\newcommand{\citeNP}[1]{\citeauthor{#1} \citeyear{#1}}

\begin{document}

\title[Multidimensional scaling for big data]{Multidimensional scaling for big data}


\author*[1,2]{\fnm{Pedro} \sur{Delicado}}\email{pedro.delicado@upc.edu}
\equalcont{These authors contributed equally to this work.}

\author[1]{\fnm{Cristian} \sur{Pach\'on-Garc\'{\i}a}}\email{cristian.pachon@upc.edu}
\equalcont{These authors contributed equally to this work.}

\affil*[1]{\orgdiv{Departament d'Estad\'{\i}stica i Investigaci\'o Operativa}, \orgname{Universitat Polit\`ecica de Catalunya$\cdot$BarcelonaTech}, \orgaddress{\street{Jordi Girona, 31}, \city{Barcelona}, \postcode{08034}, \country{Spain}}}

\affil*[2]{\orgdiv{IMTech. Institute of Mathematics of UPC-BarcelonaTech}, \orgname{Universitat Polit\`ecica de Catalunya$\cdot$BarcelonaTech}, \orgaddress{\street{Jordi Girona, 31}, \city{Barcelona}, \postcode{08034}, \country{Spain}}}




\abstract{
We present a set of algorithms implementing multidimensional scaling (MDS) for large data sets. MDS is a family of dimensionality reduction techniques using a $n \times n$ distance matrix as input, where $n$ is the number of individuals, and producing a low dimensional configuration: a $n\times r$ matrix with $r<<n$.  When $n$ is large, MDS is unaffordable with classical MDS algorithms because their extremely large memory and time requirements. 
\blue{We compare six non-standard algorithms intended to overcome these difficulties. They are} based on the central idea of partitioning the data set into small pieces, where classical MDS methods can work. 
\blue{Two of these algorithms are original proposals.}
In order to check the performance of the algorithms as well as to compare them, we  have done a simulation study. Additionally, we have used the algorithms to obtain an MDS configuration for  EMNIST: a real large data set with more than $800000$ points. \blue{We conclude that all the algorithms are appropriate to use for obtaining an MDS configuration, but we recommend to use one of our proposals, since it is a fast algorithm with satisfactory statistical properties when working with big data.} An \textsf{R} package implementing the algorithms has been created.}




\maketitle

\section{Introduction}
\label{sect:intro}
Multidimensional scaling (MDS) is a family of methods that represents 
high dimensional data in a  low dimensional space with preservation of the 
Euclidean distance between observations. 
MDS uses a $n \times n$ distance matrix as input (or, alternatively, a similarity matrix), where $n$ is the number of individuals, and producing a low dimensional configuration: a $n\times r$ matrix, where $r$ is the dimension of the low dimensional space, being much smaller than $n$.
When $n$ is large, MDS is unaffordable with classical MDS algorithms because their  extremely 
large memory ($n(n-1)/2$ values should be stored simultaneously to represent a distance 
or similarity matrix) and time requirements. The cost of the classical MDS algorithm is $O(n^3)$, 
as it requires eigendecomposition of a $n\times n$ matrix (for more details see, for instance, \citeNP{trefethen97}).

Different alternatives have been proposed in the literature, among which the following stand out: \blue{{\em FastMap} \citep{fastmap}, {\em MetricMap} \citep{metric_map}}, {\em landmark multidimensional scaling} (landmark MDS, or LMDS; \citeNP{LMDS:2004},  unpublished manuscript), \textit{fast multidimensional scaling} (fast MDS; \citeNP{Yang06afast}) and \blue{{\em pivot MDS} \citep{pivot_MDS:2007}}.

\blue{\cite{Platt:2005} shows that FastMap, MetricMap and LMDS are all 
based on a similar approximation of the eigenvectors of a large matrix, 
namely the Nystr\"om algorithm, an approximation method from Physics. 
Additionally, the author argues (based on empirical experiments) that LMDS is more accurate than FastMap and MetricMap with roughly the same computation time and can become even more accurate if allowed to be slower.
Therefore we consider LMDS in this paper, leaving aside the other two methods.}

LMDS algorithm applies first classical MDS to a subset of the data (\textit{landmark points}) and then the remaining individuals are projected onto the landmark low dimensional configuration using a distance-based triangulation procedure.
Fast MDS overcomes the problem of MDS scalability using recursive programming in combination with a \blue{data set splitting} strategy.
\blue{Pivot MDS, introduced in the literature of graph layout algorithms, is similar to LMDS but it uses the distance information between landmark and non-landmark points to improve the initial low dimensional configuration, as more relations than just those between landmark points are taken into account.}

In this work, we introduce a new non-standard MDS algorithm (\textit{divide-and-conquer MDS}) 
and an alternative form of LMDS (\textit{interpolation MDS})  which, instead of using distance-based triangulation, uses Gower's interpolation formula (\citeNP{Gower:1968}; see also the Appendix of \citeNP{GowerHand:1995}). 
Moreover, we prove that the LMDS \blue{triangulation} method proposed in 2004 coincides with the interpolation formula introduced by Gower 36 years earlier. 
Both new algorithms were proposed in \citet[Master Thesis]{Pachon_Garcia_2019}. 
\blue{In an independent work, \citeN{paradis2021} introduced {\em reduced multidimensional scaling} (reduced MDS or RMDS for short), a procedure very similar to interpolation MDS (see Section \ref{sect:interpolationMDS} below).}

\blue{In addition to these two methods, we also present \texttt{bigmds}: an \textsf{R} package \citep{Rprogram} implementing  
LMDS, 
interpolation MDS, 
RMDS, 
pivot MDS, 
divide-and-conquer MDS 
and 
fast MDS}.

The rest of the paper is organized as follows. Section \ref{sect:classical_MDS} provides a summary of classical MDS. Section \ref{sect:alg_mds} describes the \blue{six} MDS algorithms for big data considered in this paper, with particular attention to the relationship between interpolation MDS and competing methods \blue{(LMDS, pivot MDS and RMDS).} 
We also introduce the package \texttt{bigmds} implementing the \blue{six} algorithms. 
We compare them by a simulation study described in Section \ref{sect:simul}. 
In Section \ref{sect:EMNIST} we challenge \blue{all the} algorithms with a real large data set. Section \ref{sect:concl} summarizes the conclusions of the paper.

\section{Classical multidimensional scaling}
\label{sect:classical_MDS}
In this section we briefly review classical multidimensional scaling 
\blue{(\citeNP{torgerson1952multidimensional}, \citeNP{gower1966some})}.
For a more detailed explanation we refer to
\blue{Section 3.2 in \citeN{krzanowski2000principles} or} 
Chapter 12 of \citeN{BorgGroenen2005}.
Given a $n\times n$ matrix $\mathbf{\Delta}=(d_{ij}^2)$, where $d_{ij}^2$
is the squared distance between individuals $i$ and $j$, the goal of MDS is to represent the $n$ individuals in a Euclidean space with low dimensionality $r$, that is, to obtain a $n \times r$ {\em configuration matrix} $\mathbf{X}$ with orthogonal zero-mean columns such that the squared Euclidean distances between the rows of $\mathbf{X}$ are approximately equal to $\mathbf{\Delta}$. 
When equality is achieved we say that $\mathbf{X}$ is an Euclidean configuration for $\mathbf{\Delta}$. 

The columns of $\mathbf{X}$ are called \textit{principal coordinates} and they can be interpreted as the observations of $r$ latent variables for the $n$ individuals. 
Typically, the goal of MDS is dimensionality reduction, which involves looking for low dimensional configurations (that is, $r$ much lower than $n$). 

Classical MDS is one of the standard ways to obtain configuration matrices from distance matrices. For any set of $n$ vectors $\{\mathbf{y}_1,\ldots,\mathbf{y}_n\}$ in a  Euclidean space, there is a one-to-one relationship between their Euclidean distances $\{d_{ij}=\|\mathbf{y}_i - \mathbf{y}_j\|: 1\le i,j\le n\}$ and their inner products $\{q_{ij}=\mathbf{y}_i\tr \mathbf{y}_j: 1\le i,j\le n\}$: 
$d^2_{ij} = q_{ii} + q_{jj} - 2q_{ij}$ and 
$q_{ij} = -(d_{ij}^2 - d_{i.}^2 - d_{.j}^2 + d_{..}^2)/2$,
where $d_{i.}^2 = (1/n)\sum_{j = 1}^n d_{ij}^2$, 
	$d_{.j}^2 = (1/n)\sum_{i=1}^n d_{ij}^2$, and 
	$d_{..}^2 = (1/n^2)\sum_{i = 1}^n \sum_{j = 1}^n d_{ij}^2$.
	See \citeN{BorgGroenen2005} for a detailed derivation of these formulas.
In order to write the previous relationships in a matrix form, some additional definitions are convenient.  
Let $\mathbf{I}_n$ be the identity matrix of dimension $n$, and let $\mathbf{1}_n$ be the $n$-dimensional vector of ones. 
The centering matrix in dimension $n$ is defined as $\mathbf{P} = \mathbf{I}_n - \frac{1}{n} \mathbf{1}_n\mathbf{1}_n\tr$. 
The classical MDS algorithm is as follows: 

\begin{enumerate}
	\item Build the inner product matrix $\mathbf{Q} = - \frac{1}{2} \mathbf{P}
	\mathbf{\Delta} \mathbf{P}$.
	\item Obtain the eigenvalues $\lambda_i$ and eigenvectors $\mathbf{v}_i$ of $\mathbf{Q}$, 
$i=1,\ldots, n$, sorted in decreasing order of the eigenvalues. Observe that the following equality holds:
\[
\mathbf{Q} = \sum_{i=1}^{n} \lambda_i \mathbf{v}_i \mathbf{v}_i\tr.
\]
(Observe that $0$ is one of the eigenvalues of $\mathbf{Q}$, with eigenvector $\mathbf{1}_n$, because it has sum zero by rows, and that some eigenvalues may be negative when the distance matrix is not derived from a Euclidean distance measure).
	\item $\mathbf{Q}$ can be approximated by taking the $r$ greatest non-negative eigenvalues and their corresponding eigenvectors:
\[
\mathbf{Q} \approx \sum_{i=1}^{r} \lambda_i \mathbf{v}_i \mathbf{v}_i\tr = 
(\mathbf{V}_r \mathbf{\Lambda}_r^{1/2})(\mathbf{\Lambda}_r^{1/2} \mathbf{V}_r\tr ),
\]
where $\mathbf{V}_r$ has columns $\mathbf{v}_i$, $i=1,\ldots,r$, and 
$\mathbf{\Lambda}_r=\diag(\lambda_1,\ldots,\lambda_r)$.
\item Take $\mathbf{X} = \mathbf{V}_r \mathbf{\Lambda}_r^{1/2}$ 
as $r$-dimensional matrix configuration of $\mathbf{\Delta}$. 
\end{enumerate}

Observe that the configuration $\mathbf{X}$ has centered columns (because the eigenvectors of $\mathbf{Q}$ in $\mathbf{V}_r$ are orthogonal to $\mathbf{1}_n$, that is another eigenvector of $\mathbf{Q}$) with variance equal to the eigenvalues in $\mathbf{\Lambda}_r$ divided by $n$:
\begin{equation}\label{eq:Var_X} 
\mbox{Var}(\mathbf{X})
=\frac{1}{n}\mathbf{X}\tr \mathbf{X}
=\frac{1}{n} \mathbf{\Lambda}_r^{1/2} \mathbf{V}_r\tr \mathbf{V}_r \mathbf{\Lambda}_r^{1/2} 
= \frac{1}{n} \mathbf{\Lambda}_r.
\end{equation} 

The theoretical {costs} of the classical MDS algorithm {are} $\mathcal{O}(n^3)$ in time (because it requires the eigendecomposition of a $n\times n$ matrix) and $\mathcal{O}(n^2)$ in memory (because $n\times n$ matrices as $\mathbf{\Delta}$ or $\mathbf{Q}$ must be stored). These costs make classical MDS hard to deal with when the sample size is large. 

\blue{In terms of software availability, the \textsf{R} package \textsf{stats} offers a baseline functionality to compute an MDS configuration: \textsf{cmdscale}. Although we could use this implementation, we have decided to approach the computation of an MDS configuration in a faster way: use \textsf{trlan.eigen} function from \textsf{svd} package \citep{svd_package:2022} to obtain the eigendecomposition of matrix $\mathbf{Q}$ and then take the $r$ eigenvectors associated with the largest $r$ eigenvalues.
Note that the implementation presented in \textsf{svd} package uses the Lanczos eigendecomposition of a matrix (see, for instance, \citeNP{Lanczos_restart} or \citeNP{Lanczos_adaptive}), which speeds up calculations considerably.}

\blue{To compare both approaches (\textsf{stats::cmdscale} versus \textsf{svd::trlan.eigen} package implementation), we have performed 10 runs using the \textsf{microbenchmark} package \citep{microbenchmark} to measure execution times.
	For a data set of size $n=10000$,  
\textsf{stats::cmdscale} needed between 886 and 888 seconds (around 15 minutes) for each run, while \textsf{svd::trlan.eigen} took between 7.46 and 7.64 seconds in a 
computer with a processor Intel i9-10900K, 
with 64GB DDR4 of RAM memory (this is the computer used throughout this work).
We have used sample size $n=10000$ because 
the computer ran out of memory for $n= 25000$
and 
more than 20 minutes were required when using \textsf{cmdscale} for $n = 15000$.}

\section{Algorithms for multidimensional scaling with big data}
\label{sect:alg_mds}
\begin{figure}
	\begin{center}
		\begin{tabular}{ccl}
\includegraphics[scale=.25]{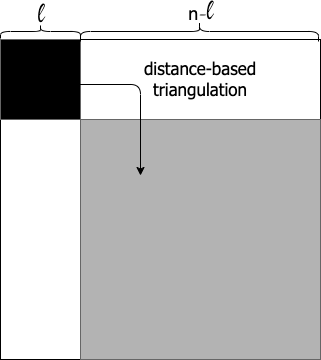}
&
\includegraphics[scale=.25]{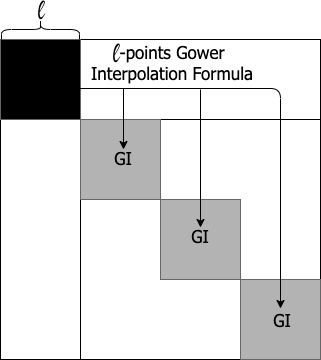}
& 
\includegraphics[scale=.25]{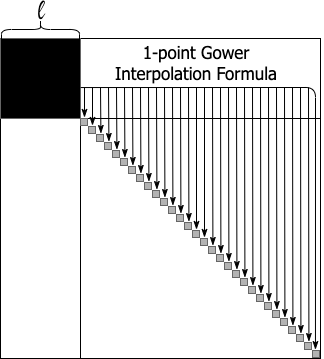}
\\[-.1cm]
{\footnotesize Landmark MDS} & 
{\footnotesize Interpolation MDS} & 
{\footnotesize \hspace*{.5cm} Reduced MDS} 
\\
\includegraphics[scale=.25]{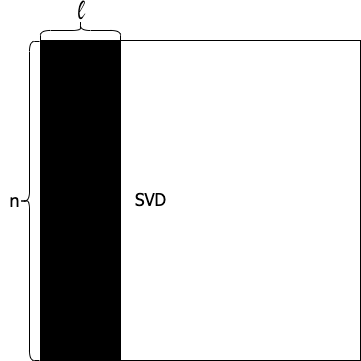}
&
\includegraphics[scale=.25]{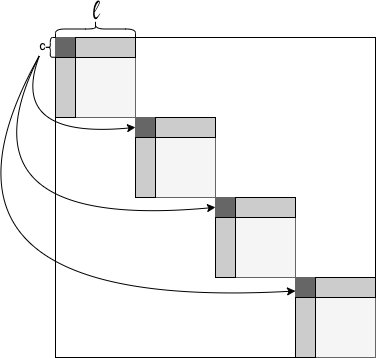}
& 
\includegraphics[scale=.25]{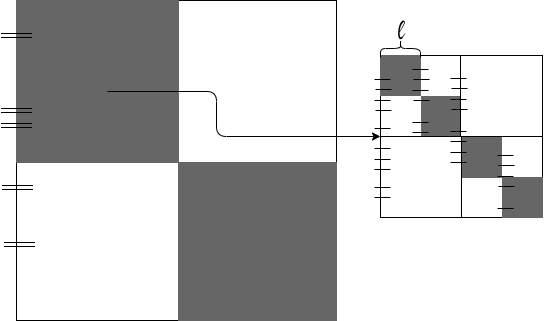}
\\[-.1cm]
{\footnotesize Pivot MDS} 
& 
{\footnotesize \hspace*{.45cm}Divide-and-conquer MDS} 
&
{\footnotesize \hspace*{.8cm} Fast MDS} \\[.1cm]
		\end{tabular} 
	\end{center}
	\caption{Schematic representation of the \blue{six} MDS algorithms described in Section \ref{sect:alg_mds}.}
	\label{fig:schema}
\end{figure}

\blue{In this section, we describe six MDS algorithms able to work with large data sets.
We start introducing interpolation MDS, which gives us the opportunity to also describe LMDS, pivot MDS and RMDS.
Then we introduce divide-and-conquer MDS and finally we talk about fast MDS. 
Figure \ref{fig:schema} schematically shows how each of these methods works.}

Given the computational cost of classical MDS algorithm, it could happen that the size of the distance matrix were so large that the computer would not be able to obtain a low dimensional configuration (we would incur in computational errors due to the capacity of the computer).  Let  $\bar{\ell}$ be the largest number for which classical MDS runs in an {\em acceptable time} (depending on both, the computer power and the user's judgment) for a distance matrix of dimension $\bar{\ell}\times \bar{\ell}$. The algorithms we are presenting require to know $\bar{\ell}$ in advance.

\subsection{Interpolation MDS}
\label{sect:interpolationMDS}
The basic idea of \blue{this} proposal is as follows.  
Given that the size of the data set is too large, we take a random sample from it of size $\ell\le \bar{\ell}$, 
to perform classical MDS to it, and to extend the obtained results to the rest of the data set by using Gower's interpolation formula (\citeNP{Gower:1968}; see also the Appendix of \citeNP{GowerHand:1995}), 
which allows us to add a new set of points to an existing MDS configuration. 
Note that this proposal follows the usual practice in Statistics when a population is too large to be examined exhaustively: 
obtaining a random sample from the population, 
analyzing it in detail and,  finally, 
extending the sampling results for the entire population.

Gower's interpolation procedure works as follows. 
Given a first data subset of size $\ell$,
let $\mathbf{D}_1=(d_{ij})$ be the $\ell \times \ell$ distance matrix between its elements, and let $\mathbf{X}_1 = \mathbf{V}_{1,r} \mathbf{\Lambda}_{1,r}^{1/2}$
be the $\ell \times r$ matrix containing its classical MDS configuration.
Consider a new data subset of size $m$ ($1\le m\le n-\ell$), and let $\mathbf{A}_{21}$ be the $m \times \ell$ distance matrix between its $m$ elements and the $\ell$ ones in the first data subset. 
One wants to project these new $m$ elements into the existing MDS configuration in such a way that the Euclidean distances between the new projected points and the original ones are as close as possible to the elements of $\mathbf{A}_{21}$. 
We briefly summarize how to do so using Gower's interpolation formula.
Define $\mathbf{Q}_1 = - \frac{1}{2} \mathbf{P} \mathbf{\Delta}_1 \mathbf{P}\tr $, 
where $\mathbf{\Delta}_1 = (d_{ij}^2)$ and $\mathbf{P} = \mathbf{I}_{\ell} - \frac{1}{\ell}\mathbf{1}_{\ell}\mathbf{1}_{\ell}\tr $.
Let $\mathbf{q}_1$ be the diagonal of $\mathbf{Q}_1$, treated as a column vector. 
Let $\mathbf{A}_{21}^2$ be the matrix of the square of the elements of $\mathbf{A}_{21}$.
Let $\mathbf{S}_1$ be the variance-covariance matrix of the $r$ columns of $\mathbf{X}_1$. 
So,  by equation (\ref{eq:Var_X}), $\mathbf{S}_1=(1/\ell)\mathbf{\Lambda}_{1,r}$.
Gower's interpolation formula, equation (8) in \citeN{Gower:1968}, states that the interpolated coordinates for the new $m$ observations are given by
\begin{equation} \label{gower_f}
	\mathbf{\dot {X}}_2 = 
\frac{1}{2\ell} (\mathbf{1}_m\mathbf{q}_1\tr  - \mathbf{A}_{21}^2) \mathbf{X}_1\mathbf{S}_1^{-1}
.
\end{equation}

Observe that Gower's interpolation formula is valid for any number $m\ge 1$ of elements in the second data subset. 
Nevertheless, for large values of $m$ the memory limitations reported for classical MDS could appear here because formula (\ref{gower_f}) involves matrices of dimension $m\times \ell$. 
Therefore we propose to use $m=\ell$ when projecting new observations into an existing MDS configuration.

The proposed interpolation MDS algorithm operates as follows. 
First, the data set of size $n$ is divided into $p=\lceil n/\ell\rceil$ parts.
The first data subset is used to compute $\mathbf{X}_1$ and the other elements with Gower's interpolation formula (\ref{gower_f}). Then, we use this formula to obtain 
$\mathbf{\dot {X}}_j$, where $j \in \{2, \dots, p\}$.

Finally, all the interpolated partial MDS configurations are concatenated by rows to obtain the global $n\times r$ configuration:
\[
\mathbf{X}= \left[\mathbf{X}_1\tr\mid  \mathbf{\dot {X}}_2\tr\mid \cdots\mid 
\mathbf{\dot {X}}_p\tr \right]\tr.
\]
The eigenvalues $\lambda_i^1$, $i=1,\ldots,r$, obtained when applying classical MDS to the first data subset, divided by $\ell$, are estimations of the variance of the columns of $\mathbf{X}$.

Observe that if the number of rows of the original data set is such that 
it allows to run classical MDS over the whole data set, 
then $p=1$ and interpolation MDS is just the classical MDS. 

The most costly operation in this algorithm is the computation of the distance matrix $\mathbf{A}_{21}$, that in our case is of order $\mathcal{O}(\ell^2)$ because we use $m=\ell$. 
This operation is repeated $p$ times, with $p \approx n/\ell$. 
So the computation cost in time of this algorithm is  $\mathcal{O}(n\ell)$.

Interpolation MDS shares similarities with \blue{three} existing algorithms: LMDS, \blue{pivot MDS} and RMDS, \blue{all three with computation cost $\mathcal{O}(n\ell)$ in time}
(we are grateful to a referee who pointed us to these methods).
In interpolation MDS, LMDS and RMDS, a data subset is selected (called set of landmark points in LMDS) and classical MDS is performed on it to obtain the corresponding low dimensional configuration.
Then the reaming points are projected onto this initial configuration.
The three methods use different projection strategies at this second step.

On the one hand, interpolation MDS and RMDS use Gower's interpolation formula (\ref{gower_f}):
interpolation MDS uses a matrix version of that formula (just as stated in equation \ref{gower_f}) which is valid for interpolating simultaneously a large number of points not used in the initial MDS,  
whereas RMDS uses the version of Gower’s formula valid only for one new point and it needs to visit all the points not used in the initial MDS, one at a time. 

On the other hand, LMDS uses a distance-based triangulation procedure to project the non-landmark individuals. 
At a first glance, this procedure is different from Gower's interpolation formula. 
Nevertheless, when the LMDS projection formula is carefully examined, it can be seen that it coincides in fact with Gower's interpolation formula (\ref{gower_f}). 
This is a result not previously reported in the literature, as far as we know. 
Indeed, following equation (3) in \citeN{LMDS:2004}, LMDS projection formula can be written as follows:
\begin{equation} \label{eq:LMDS_proj}
	\blue{
	\mathbf{X}_2^{\mathrm{\footnotesize{LMDS}}} = }
\frac{1}{2} 
	(\mathbf{1}_m\mathbf{d}_1\tr  - \mathbf{A}_{21}^2) 
\mathbf{V}_{1,r} \mathbf{\Lambda}_{1,r}^{-1/2}
\end{equation}
where $\mathbf{d}_1=(1/\ell)\mathbf{\Delta}_1 \mathbf{1}_{\ell}$ is the vector of average squared distances between each landmark point and the other. 
Similarities between formulas (\ref{gower_f}) and (\ref{eq:LMDS_proj}) are evident. 
Moreover, observe that
\[
\mathbf{q}_1
=\mathrm{Diag}(\mathbf{Q}_1)
=-\frac{1}{2}\mathrm{Diag}\left(\mathbf{P} \mathbf{\Delta}_1 \mathbf{P}\tr\right)
=
\]
\[
-\frac{1}{2}\mathrm{Diag}\left(\mathbf{\Delta}_1 
- \mathbf{d}_1 \mathbf{1}_{\ell}\tr  
- \mathbf{1}_{\ell} \mathbf{d}_1\tr  
+ \mathbf{1}_{\ell} \mathbf{1}_{\ell}\tr \bar{\mathbf{d}}_1
\right)
=
\]
\[
-\frac{1}{2}\left(\mathbf{0}_{\ell} 
- \mathbf{d}_1
- \mathbf{d}_1
+ \mathbf{1}_{\ell} \bar{\mathbf{d}}_1
\right)
=
\mathbf{d}_1 - \frac{\bar{\mathbf{d}}_1}{2} \mathbf{1}_{\ell},
\]
where $\bar{\mathbf{d}}_1$ is the average of the square distance values in $\mathbf{\Delta}_1$.
\blue{Additionally, 
\[
\frac{1}{\ell}
\mathbf{X}_1\mathbf{S}_1^{-1}=
\frac{1}{\ell}
\left(\mathbf{V}_{1,r} \mathbf{\Lambda}_{1,r}^{1/2}\right)
\left((1/\ell)\mathbf{\Lambda}_{1,r}\right)^{-1}=
\mathbf{V}_{1,r} \mathbf{\Lambda}_{1,r}^{-1/2}.
\]}
Therefore,
\[
\mathbf{X}_2^{\mathrm{\footnotesize{LMDS}}} = 
\frac{1}{2\ell} (\mathbf{1}_m \mathbf{q}_1\tr + \frac{\bar{\mathbf{d}}_1}{2} \mathbf{1}_m \mathbf{1}_{\ell}\tr - \mathbf{A}_{21}^2) \mathbf{X}_1\mathbf{S}_1^{-1}
=
\mathbf{\dot {X}}_2 + 
\frac{\bar{\mathbf{d}}_1}{4\ell} \mathbf{1}_m \mathbf{1}_{\ell}\tr \mathbf{X}_1\mathbf{S}_1^{-1} 
= \mathbf{\dot {X}}_2,
\]
and we conclude that LMDS projection coincides with Gower's interpolation formula.
We have used that $\mathbf{X}_1$ has zero mean by columns in the last step. 
So we have proved the following Proposition. 
\begin{proposition}\label{prop:LMDS_Gower}
Distance-based triangulation procedure used in LMDS coincides with Gower's interpolation formula.
\end{proposition}

We have seen that interpolation MDS, LMDS and RMDS are essentially three variations of the same procedure.
Nevertheless, they differ in the way the initial data subset is selected:
interpolation MDS chooses it at random, 
LMDS uses a MaxMin greedy optimization procedure, 
and RMDS follows a set of heuristic rules (already used in \citeNP{Paradis:2018}) intended to ensure the inclusion of both central and peripheral observations.
Note that random selection is also an option in the available implementations of LMDS and RMDS. 
In their second step, the three algorithms use Gower's formula for projecting the remaining points, but there are some subtle differences between them:  
RMDS projects each point at a time, 
interpolation MDS performs this operation in blocs of $\ell$ points, 
and LMDS projects all points in a single step.

\blue{
Finally, pivot MDS is an approximation of classical MDS with a similar approach to LMDS. 
Once the subset of $\ell$ landmark points have been selected (in this context, they are called {\em pivot points}), let $\mathbf{C}$ be the $n\times \ell$ submatrix of $\mathbf{Q}$ containing the inner products between the pivot points and all the points in the data set. 
The singular value decomposition of $\mathbf{C}$ is used to approximate that of $\mathbf{Q}$, 
whose $r$ first eigenvectors lead to the pivot MDS low dimensional configuration.
Recall that LMDS is based on the eigendecomposition of the  $\ell\times \ell$ submatrix of $\mathbf{Q}$ containing only inner products of landmark points. 
}

\subsection{Divide-and-conquer MDS}
\label{sect:divideMDS}

We base this algorithm on the principle of dividing and conquering. Roughly speaking, a large data set is divided into parts, then MDS is performed over every part and, finally, the partial configurations are combined so that all the points lie on the same coordinate system. Let us go into the details.

Let $n$ be the number of individuals of the original data set, which is divided into $p$ parts of size $\ell$, where $\ell\le \bar{\ell}$.
The algorithm requires that all the partitions have $c$ individuals in common. 
Those $c$ individuals are used in order to connect the MDS partial configuration obtained from each part and we name them {\em connecting points}. 
This number $c$ should be large enough to guarantee good links between partial configurations, but as small as possible to favor efficient computations.
Given that the partial configurations will be connected by a Procrustes transformation (see, for instance, Chapter 20 of \citeNP{BorgGroenen2005}), $c$ must be at least equal to $r+1$ (to avoid reflections), where $r$ is the required low dimension we are looking for when applying classical MDS to every part of the data set.

The divide-and-conquer MDS starts selecting at random the $c$ connecting points from the data set (selection strategies different from randomness could be used, \blue{as done in LMDS or RMDS}; see Section \ref{sect:interpolationMDS}).
%
%
Then, $\mathbf{X}$ is divided into $p$ data subsets, where $p=\lceil 1 + (n-\ell)/(\ell-c)\rceil$ is the lowest integer larger than or equal to $1 + (n-\ell)/(\ell-c)$.  These data subsets are defined containing the $c$ connecting points plus $\ell-c$ randomly selected (without replacement) points from the remaining $n-c$.
Classical MDS is applied to each data subset, with configurations of dimension $r$. 
Let $\mathbf{X}_j$, $j=1,\ldots,p$, be the $\ell\times r$ configuration obtained from the $j$-th data subset. 

Since all the partitions share $c$ points, the first configuration $\mathbf{X}_1$ can be aligned with any other $\mathbf{X}_j$, $j\ge 2$, using a Procrustes transformation. 
In order to do that, 
let $\mathbf{X}_1^c$ and $\mathbf{X}_j^c$ be the $c\times r$ matrices corresponding to the connecting points in $\mathbf{X}_1$ and $\mathbf{X}_j$ respectively. 
The Procrustes procedure is applied to $\mathbf{X}_1^c$ and $\mathbf{X}_j^c$ and the parameters $\mathbf{T}_j\in \mathbb{R}^{r\times r}$ and  $\mathbf{t}_j\in \mathbb{R}^{r}$ are 
obtained  so that 
\[
\mathbf{X}_1^c \approx \mathbf{X}_j^c \mathbf{T}_j + \mathbf{1}_c \mathbf{t}_j\tr.
\]
Let $\mathbf{X}_j^a$ be $\mathbf{X}_j$ without the connecting $c$ points,
and let
\[
\mathbf{\ddot {X}}_j = \mathbf{X}_j^a \mathbf{T}_j + \mathbf{1}_{l-c} \mathbf{t}_j \tr
\]
be the $(\ell-c)\times r$ matrix with the $j$-configuration (excluding the connecting points) aligned with respect to $\mathbf{X}_1$.
Finally, all the aligned partial MDS configurations are concatenated by rows to obtain the global $n\times r$ configuration:
\[
\mathbf{X}= \left[\mathbf{X}_1\tr \mid  \mathbf{\ddot {X}}_2\tr\mid \cdots\mid 
\mathbf{\ddot {X}}_p\tr \right]\tr.
\]

When classical MDS is applied to each data subset, in addition to the $r$-dimensional configuration $\mathbf{X}_j$, it provides the 
eigenvalues $\lambda_i^j$, $i=1,\ldots,r$, of the inner product matrix $\mathbf{Q}_j$ which, divided by the size of data subset, coincide with the eigenvalues of the variance matrix of the columns of $\mathbf{X}_j$, shared as well by $\mathbf{\ddot{X}}_j$ because $\mathbf{T}_j$ is an orthogonal matrix. \blue{Therefore, we can define a set of estimators for the first $r$ eigenvalues as }
\[
\bar{\lambda}_i = \frac{1}{p}\sum_{j =1}^p \frac{\lambda_i^j}{n_j}, \, i=1,\ldots, r,
\]
\blue{where $n_j$ is the size of the $j$-th data subset ($n_j=\ell$ for all $j$, except perhaps for the last one).} Observe that $\bar{\lambda}_i$ is also an estimation of the variance of the $i$-th column in final MDS configuration $\mathbf{X}$.

In terms of computation time, the most costly operation is to obtain an MDS configuration for an $\ell \times \ell$ matrix, which cost is $\mathcal{O}(\ell^3)$. This operation is performed $p$ times, being $p \approx n/\ell$. Therefore, the total cost is  $\mathcal{O}(n\ell^2)$.
As in the previous algorithm, note that if $n \leq \ell$ then $p=1$ and divide-and-conquer MDS is just classical MDS.

\subsection{Fast MDS}
\label{sect:fastMDS}

As in the previous approaches, fast MDS also randomly divides the whole sample data set of size $n$ into several data subsets, but now the size of the data subsets can be larger than $\ell$ (with $\ell\le \bar{\ell}$) because of the recursive strategy: 
fast MDS is applied again when the {sizes of the data} subsets are larger than $\ell$.
\citeN{Yang06afast} do not give precise indications for choosing $\ell$: they say that $\ell$ must be the size of {\em the largest matrix that allows MDS to be executed efficiently}, from which it follows that $\ell\le \bar{\ell}$.

In the last step of the fast MDS algorithm, the partial MDS configurations obtained for each data subset are combined into a global MDS configuration by a Procrustes transformation (as in divide-and-conquer MDS, Section \ref{sect:divideMDS}). 
To do so, a small subset of size $s$ is randomly selected from each data subset (\citeNP{Yang06afast} call them the {\em sampling points}).
The role of $s$ in fast MDS is equivalent to that of $c$ (the amount of connecting points) in divide-and-conquer MDS, and the same considerations for its choice apply here. In particular, $s\ge r+1$.

The selected sampling points from each data subset are joined to form an {\em alignment set}, over which classical MDS is performed giving rise to an {\em alignment configuration}: a $\ell\times r$ matrix $\mathbf{X}_{\textrm{align}}$. 
In order to be able to apply classical MDS to the alignment set, its size must not exceed the limit size $\ell$. Therefore, the number $p$ of data subsets is taken as $p=\lfloor\ell/s\rfloor$ (the integer part of $\ell/s$).  

Each one of the $p$ data subsets has size $\tilde{n} = \lceil n/p \rceil$ (except perhaps the last one). If $\tilde{n}  \leq \ell$ then classical MDS is applied to each data subset. Otherwise, fast MDS is recursively applied.
In either case, a final MDS configuration is obtained for each data subset,
namely the $\tilde{n}\times r$ matrices $\mathbf{X}_j, j=1,\ldots, p$.

Every data subset shares $s$ points with the alignment set. 
Therefore every MDS configuration $\mathbf{X}_j$, $j\ge 1$, can be aligned with the alignment configuration $\mathbf{X}_{\textrm{align}}$ using a Procrustes transformation. 
Let $\mathbf{X}_{\textrm{align},j}^s$ and $\mathbf{X}_j^s$ be the $s\times r$ matrices corresponding to the $j$-th set of sampling points in $\mathbf{X}_{\textrm{align}}$ and $\mathbf{X}_j$ respectively. 
The Procrustes procedure is applied to $\mathbf{X}_{\textrm{align},j}^s$ and $\mathbf{X}_j^s$ and the parameters, $\mathbf{T}_j\in \mathbb{R}^{r\times r}$ and  $\mathbf{t}_j\in \mathbb{R}^{r}$ are 
obtained,  so that 
\[
\mathbf{X}_{\textrm{align},j}^s\approx \mathbf{X}_j^s \mathbf{T}_j + \mathbf{1}_s \mathbf{t}_j\tr.
\]
Let 
\[
\mathbf{\tilde{X}}_j = \mathbf{X}_j \mathbf{T}_j + \mathbf{1}_{\tilde{n}} \mathbf{t}_j \tr
\]
be the $\tilde{n}\times r$ matrix with the $j$-configuration  aligned with respect to $\mathbf{X}_{\textrm{align}}$.
Finally, all the aligned partial MDS configurations are concatenated by rows to obtain the global $n\times r$ configuration:
\[
\mathbf{X}= \left[\mathbf{\tilde{X}}_1\tr\mid  \mathbf{\tilde{X}}_2\tr\mid \cdots\mid 
\mathbf{\tilde{X}}_p\tr \right]\tr.
\]

As in the previous algorithms, note that if $n \leq \ell$ then $p=1$ and 
fast MDS is just classical MDS.
Average of eigenvalues are defined as in divide-and-conquer MDS. 

\citeN{Yang06afast} use the Master theorem for recurrent algorithms (\citeNP{bentley1980general}) to establish that the computation cost in time of the fast MDS algorithm is 
$\mathcal{O}(n\log n)$.

At a first sight divide-and-conquer MDS and fast MDS show some similarities (classical MDS is applied to small portions of the data, and then the pieces are joined by Procrustes transformation).
Nevertheless, they are significantly different mainly because fast MDS is a recursive algorithm while divide-and-conquer MDS is not.
This difference implies, for instance, that in divide-and-conquer MDS the data subsets at which the classical MDS is applied have always the same size $\ell$ (controlled by the user),
whereas in fast MDS these sizes can not be fixed in advance (they depend on the successive recursive partitioning process).
This difference has practical performance implications, as it can be seen in Section \ref{sect:simul}.

\subsection{\textsf{bigmds}: the \textsf{R} package to do MDS with big data}
\label{sect:bigmds_package}

In order to make these methods available, we have published an \textsf{R} package in \textsf{CRAN}: \url{https://cran.r-project.org/web/packages/bigmds}.
The core of the package consists of \blue{six} methods:
\blue{\textsf{landmark\_mds},}
\textsf{interpolation\_mds},
\blue{\textsf{reduced\_mds},}
\blue{\textsf{pivot\_mds},}
\textsf{divide\_and\_conquer\_mds} 
and 
\textsf{fast\_mds}. 
Each of these functions provides an MDS configuration following the procedures described in section \ref{sect:alg_mds}.
We also developed a Procrustes function which is used by  
\blue{\textsf{divide\_and\_conquer\_mds} 
and 
\textsf{fast\_mds}}. 
We {followed} \citeN{BorgGroenen2005} in order to obtain Procrustes parameters.
The package has also a development version which is 
available in \textsf{GitHub}: 
\url{https://github.com/pachoning/bigmds}. Finally, as a classical MDS algorithm we used \blue{\textsf{trlan.eigen} function (from \textsf{svd} package)  to perform the eigendecomposition and then the desirable number of columns is taken from the matrix that contains the eigenvectors.}

\section{Simulation study}
\label{sect:simul}
In this section, we present a simulation study to evaluate the \blue{six} MDS methods.
In particular, we address the following questions: 
(1) the ability to capture the right data dimensionality and 
(2) the speed of the algorithms.
We have not included the full classical MDS in our simulation study because it can not run for sample sizes greater than \blue{25000} (as we report in Section \ref{sect:classical_MDS}). 

\subsection{Design of the simulation}
\label{sect:design}
Different experiments {were} conducted in order to answer the previous questions. 
At each experiment, data matrices $\mathcal{Y}$ of dimension $n \times k$ {were} generated, which rows {were} considered to be the individuals in the data set.
Euclidean distances between rows of $\mathcal{Y}$ {were} used throughout the study.
Several scenarios {were} explored, taking into account different factors:

\begin{description}
	\item[{Sample size.}] Different sample sizes  $n$ {were} taken into account, combining small
	data sets and large ones. A total of \blue{eight} sample sizes {were} used: $5 000$, $10 000$, $20 000$, $100 000$, \blue{$250 000$, $500 000$, $750 000$} and $1 000 000$.
	
	\item[{Data dimension.}] The considered number of columns $k$ {were} 10 and 100.
	
	\item[{Dominant dimension.}] The first $h$ columns in $\mathcal{Y}$ {had} variance equal to 15, while the other $k-h$ {had} variance equal to $1$. 
	{Throughout this paper we refer} to $h$ as the \textit{dominant dimension}. The idea of this {was} to see if the algorithms {were} able to capture the relevant data dimensionality.
	We considered two values for \blue{$h$: 2 and 10}.
\end{description}

There was a total of  \blue{32} scenarios to simulate (\blue{8} sample sizes, 2 data dimensions, and \blue{2} dominant dimensions). Each scenario was replicated 100 times. So, a total of \blue{3200} simulations were carried out. 

For every simulation, the data matrix $\mathcal{Y}$ was generated from a multivariate normal distribution with zero mean independent coordinates and variances $15$ for the first $h$ columns and {variance} $1$ for the others.
The \blue{six} MDS algorithms {were} run based on Euclidean distances between rows of $\mathcal{Y}$.
All the algorithms were executed requiring as many columns $r$ as the dominant dimension of the simulated data set, i.e, $r=h$. 
Therefore, the resulting low dimensional MDS configurations $\mathbf{X}$ {had} dimension $n\times h$.	
In addition to $\mathbf{X}$, the elapsed time {was} stored for each simulation. 

Note that the original data set, $\mathcal{Y}$, {was} already an MDS configuration by construction, since we {simulated} independent columns with zero mean. 
Therefore, even though $n$ {was} so large that classical MDS {could} not be calculated,
the first $h$ columns of $\mathcal{Y}$ {could} be taken as a benchmark classical MDS solution to which compare against the MDS configurations $\mathbf{X}$ provided by the \blue{six} algorithms.

In order to test the quality of the algorithms as well as 
the time needed to compute the MDS configurations, some metrics {were} calculated:

\begin{itemize}
	\item The quality of the results {was} measured by the following statistics:
	\begin{itemize}
		\item Correlation between the dominant directions of the data and the  corresponding dimensions provided by the algorithms.
		Note that any rotation of an MDS configuration $\mathbf{X}$ {would have led} to another equally valid configuration. Therefore, before computing the correlations between the first $h$ columns of $\mathcal{Y}$, which we denote by  $\mathbf{\mathcal{Y}^h}$, and those of $\mathbf{X}$, we {had to} be sure that both matrices were {\em correctly aligned}, in the sense that we {were} using the rotation of $\mathbf{X}$ that best {fitted} the columns of $\mathcal{Y}$. A Procrustes transformation {was} done to achieve this alignment. 
		\\
		It is worth mentioning that alternative quality measurements are defined directly comparing distances between the original data and those computed from the obtained configurations (see, for instance, Chapter 11 in  \citeNP{BorgGroenen2005}). Nevertheless, when the sample size is moderate or large, it is not possible to compute all the distances between individuals. So it is impossible to use distances to compare solutions.
		\item {\em Bias} and {\em Root Mean Squared Error} (RMSE) of the eigenvalues $\bar{\lambda}_i$, $i=1,\ldots,h$, as estimators of the variance of the first $h$ columns of $\mathbf{X}$ (namely, $15$).
	\end{itemize}
	\item The computational efficiency {was} measured by the average time to get the MDS configurations over the $100$ replications of each scenario. Specifically, we measured the elapsed time between start and finish of each algorithm.
\end{itemize}

The \blue{six} algorithms require to specify a value for $\ell$ parameter. 
Section \ref{sect:l_election} below, entirely devoted to the choice of $\ell$, justifies the following values: 
\blue{
	$\ell=250$ for LMDS, interpolaton MDS and RMDS,
	$\ell=200$ for pivot MDS, 
	$\ell=400$ for divide-and-conquer MDS, 
	and $\ell=600$ fast MDS.}

Divide-and-conquer MDS and fast MDS have an additional parameter each: 
the number $c$ of connecting points in divide-and-conquer MDS, and the number $s$ of sampling points in fast MDS. 
Both $c$ and $s$ must be greater than or equal to the number $r$ of columns required for the MDS configuration. 
\citeN{Yang06afast} {used} $s=2r$,
\blue{but we have chosen more conservative values:  $s=c=5r$.}
Using lower values for $c$ or $s$ could {have led}  to incorrect MDS configurations, as there would {have been} very few points on which to base the Procrustes transformations.
On the other hand, using larger values for $c$ or $s$ would {have lengthened} the time of the algorithms.

\subsection{Choosing the partition size $\ell$}
\label{sect:l_election}
Before running the complete simulation study, we {examined} the effect of the {partition} size $\ell$ on the algorithms efficiency, 
measuring the ability to recover the low dimensional data structure (which we quantified in two different ways, as explained in Section \ref{sect:design}), 
and the computation time. 
We ran a simple experiment using 20 simulated data sets that {were} analyzed with the \blue{six} algorithms using \blue{9 different values of $\ell$: from 100 to 300 in steps of 50, and from 400 to 700 in steps of 100.} 
A data matrix $\mathcal{Y}$ of dimension $1 000 000 \times 100$ {was} generated, with elements being independent random normal observations with zero mean  and variance equal to 15 for the first 10 columns, {and variance equal to 1 for the remaining ones.} 
The \blue{six} algorithms {were} executed with $r=10$, and $s=c=\blue{50}$.

\blue{Regarding the range of $\ell$, we tried to go further than 700 but LMDS and pivot MDS experienced memory problems. This was due to the nature of these algorithms (a distance matrix of size $\ell\times n$ must be stored). Even though the number $\ell$ of landmark or pivot points is small, the sample size $n$ may be large enough so that the rectangular distance matrix can not fit in memory.}

Each algorithm {was} evaluated for every value of $\ell$ with three performance measures: 
the correlation of the 10 first columns of $\mathcal{Y}$ with the obtained configuration matrix,
the proximity of the $r=10$ estimated eigenvalues to 15 (their theoretical value),
and the computation time (in seconds). 
More details on these measures are given in Sections \ref{sect:simul.corr},
\ref{sect:simul.eigenvals} and
\ref{sect:simul.time}, respectively. In particular, we avoided rotation, reflection, and translation problems by performing a Procrustes transformation to the different MDS configurations in order to align them to the original data set before computing correlations.

\begin{figure}
	\begin{subfigure}{.5\textwidth}
		\centering
		\includegraphics[scale=0.55]{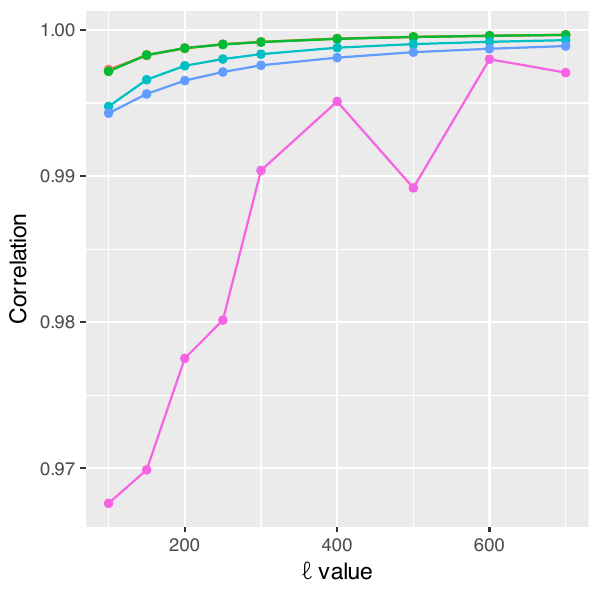}  
		\caption{}
		\label{fig:sub_corr}
	\end{subfigure}
	\begin{subfigure}{.5\textwidth}
		\centering
		\includegraphics[scale=0.55]{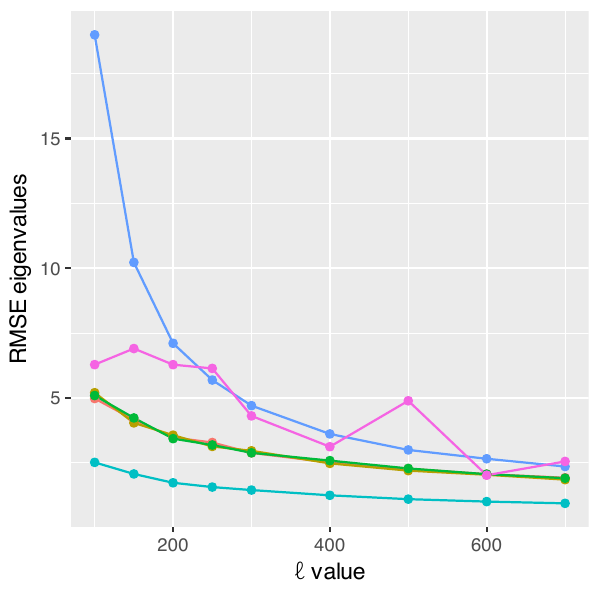}  
		\caption{}
		\label{fig:sub_bias}
	\end{subfigure}
	\begin{subfigure}{.5\textwidth}
		\centering
		\includegraphics[scale=0.55]{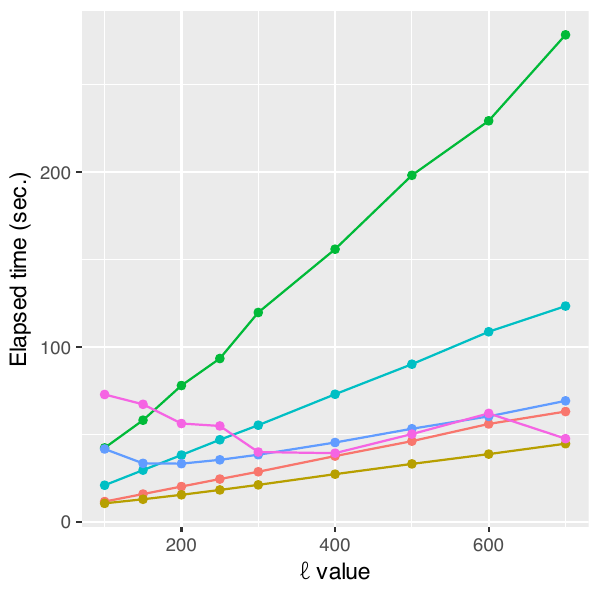}  
		\caption{}
		\label{fig:sub_time}
	\end{subfigure}
	\begin{subfigure}{.5\textwidth}
	\includegraphics[scale=.55]{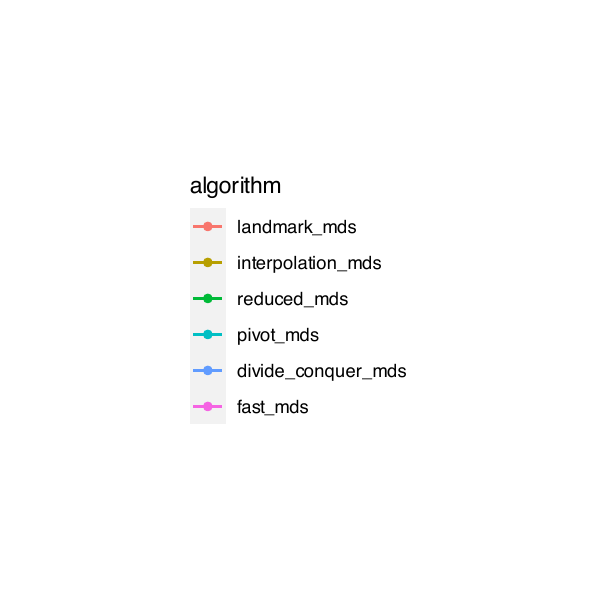}  
\end{subfigure}
	\caption{
		(\ref{fig:sub_corr}) Mean correlation coefficient for the 10 first columns. 
		Note that all correlation coefficients were above $0.9$.
		(\ref{fig:sub_bias}) Root Mean Squared Error of $\lambda_i$, $i \in \{1, \dots, 10\}$, as estimators of their theoretical value $15$.
		(\ref{fig:sub_time}) Mean time required to obtain an MDS configuration. 
	}
	\label{fig:ell_performance}
\end{figure}

Figure \ref{fig:ell_performance} shows the performance measures for each algorithm.
We looked for the values of $\ell$ at which a compromise between the three criteria was  achieved.
\blue{Observe that the results in correlation and RMSE for LMDS, interpolation MDS and RMDS were almost indistinguishable, while elapsed times were different (interpolation MDS was faster than LMDS, which was faster than RMDS).}

We considered that satisfactory results were met by choosing \blue{
	$\ell=250$ for LMDS, interpolation MDS and RMDS (we decided to use the same value of $\ell$ for these methods, given their common performance),
	$\ell=200$ for pivot MDS, 
	$\ell=400$ for divide-and-conquer MDS, and 
	$\ell=600$ fast MDS. 
}

\blue{
	Notice that the three performance measures for the fast MDS algorithm depend on $\ell$ in a non-monotonous way,
	possibly due to successive recursive divisions. 
	For instance, it can be checked that for $n= 1 000 000$ and $\ell=500$ the algorithm required a total of $10 000$ partitions, with an average size of $100$ points, while when $\ell=600$ the number of partitions {was} $1 730$ with average size $578$. 
	The small size of partitions was probably the reason of the poor behavior  in correlation and RMSE for $\ell=500$.
	Observe that the values of $\ell$ leading to small partitions may be different for other sample sizes.
	Therefore our {decision} of using $\ell=600$ for fast MDS {was} appropriate for $n= 1 000 000$ but could {have not been} the best choice for other values of $n$.
	In this sense, the {chosen} values of $\ell$ {were} less robust against changes in $n$ for fast MDS than for the other MDS algorithms.
}

\subsection{Results on correlation with the dominant directions}\label{sect:simul.corr}
This section is aimed to study the ability of the \blue{six} MDS algorithms to capture the dominant directions. 
Given a simulated data set, $\mathcal{Y}$, there were \blue{seven} MDS configurations related to the data set: 
the data set itself,  $\mathcal{Y}$, and one per each of the \blue{six} methods {proposed}. 
After applying Procrustes to a given MDS configuration, $\mathbf{X}$, the columns of the resulting matrix should {have been} highly correlated with the dominant directions of $\mathcal{Y}$ (as described  in Section \ref{sect:design}).

Table \ref{table:correlation} contains the $2.5\%$ quantile ($\q_{0.025}$), the mean value ($\mean$) and the $97.5\%$ quantile ($\q_{0.975}$)  for the correlation coefficients for each of the \blue{six} algorithms. 
For each scenario described in Section \ref{sect:design}, a total of $h$ correlation coefficients were computed, where $h$ was the dominant dimension of the scenario. 
Then, \blue{192} correlation coefficients {were} derived from a single replication of the \blue{32} scenarios (\blue{$192=8 \times 2 \times (2+10)$}). 
As performed 100 replications, Table \ref{table:correlation} shows descriptive statistics of correlation coefficients sets of size \blue{$19200$}.
It can be seen that there was a high correlation between the MDS configurations and the dominant directions of $\mathcal{Y}$ for each of the 
\blue{six} algorithms. 
Furthermore, \blue{LMDS}, interpolation MDS and \blue{RMDS} were the algorithms that provided the MDS configurations most correlated with the dominant directions, followed by divide-and-conquer MDS, \blue{pivot MDS} and then fast MDS.

\begin{table}
	\caption{
		Quantiles of order $2.5\%$ $(\q_{0.025})$ and  $97.5\%$ $(\q_{0.975})$, and $\mean$ values for the correlation coefficients between the original variables and the ones recovered by the \blue{six} MDS methods.
	}
	\label{table:correlation}
	\centering
	\blue{
	\begin{tabular}{lrrr}
	\hline
	algorithm & $\q_{0.025}$ & mean & $\q_{0.975}$ \\
	\hline
	LMDS                   & 0.99869 & 0.99950 & 1 \\	
	Interpolation MDS      & 0.99868 & 0.99949 & 1 \\
	RMDS                   & 0.99868 & 0.99949 & 1 \\
	Pivot MDS              & 0.99621 & 0.99824 & 0.99988 \\
	Divide-and-conquer MDS & 0.99774 &0.99845 & 0.99915 \\
    Fast MDS               & 0.98278 & 0.99417 & 0.99886\\
	\hline
	\end{tabular}}
\end{table}

\subsection{Results on eigenvalues}\label{sect:simul.eigenvals}
\begin{figure}
	\centering
	\includegraphics[scale=.9]{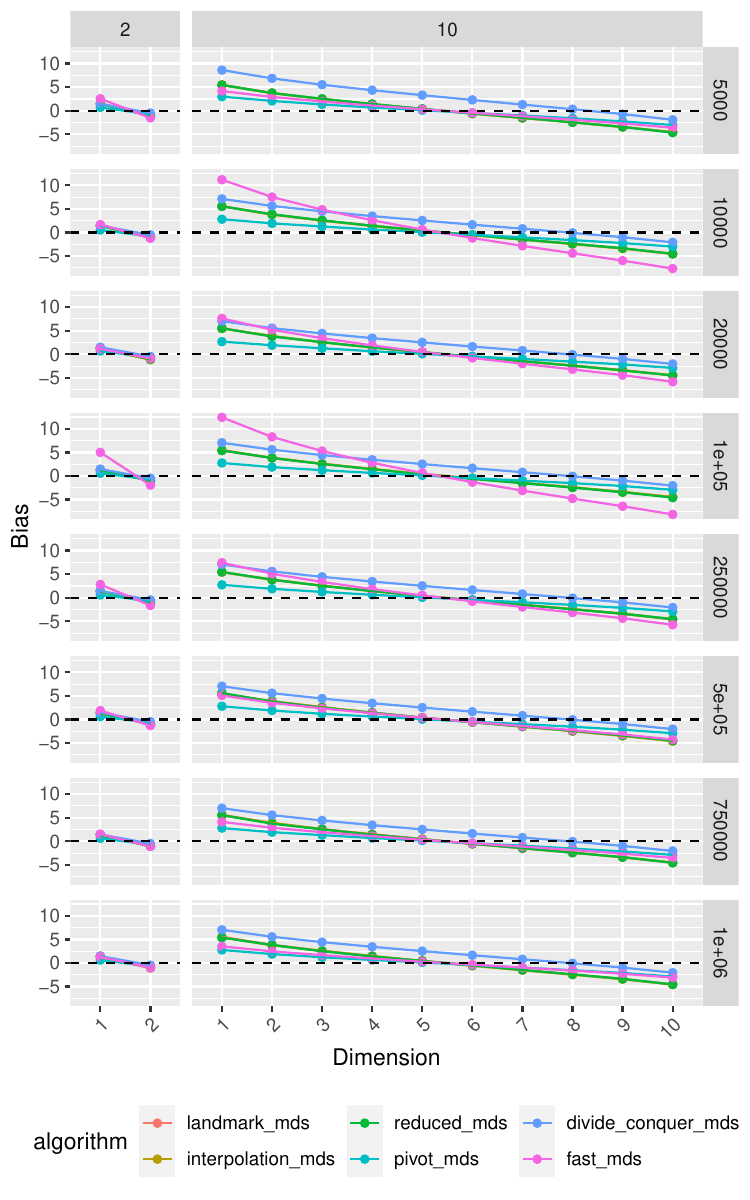}
	\caption{
		Bias of the estimators for the variance of the dominant directions.
	}
	\label{fig:bias}
\end{figure}

\begin{figure}
	\centering
	\includegraphics[scale=.9]{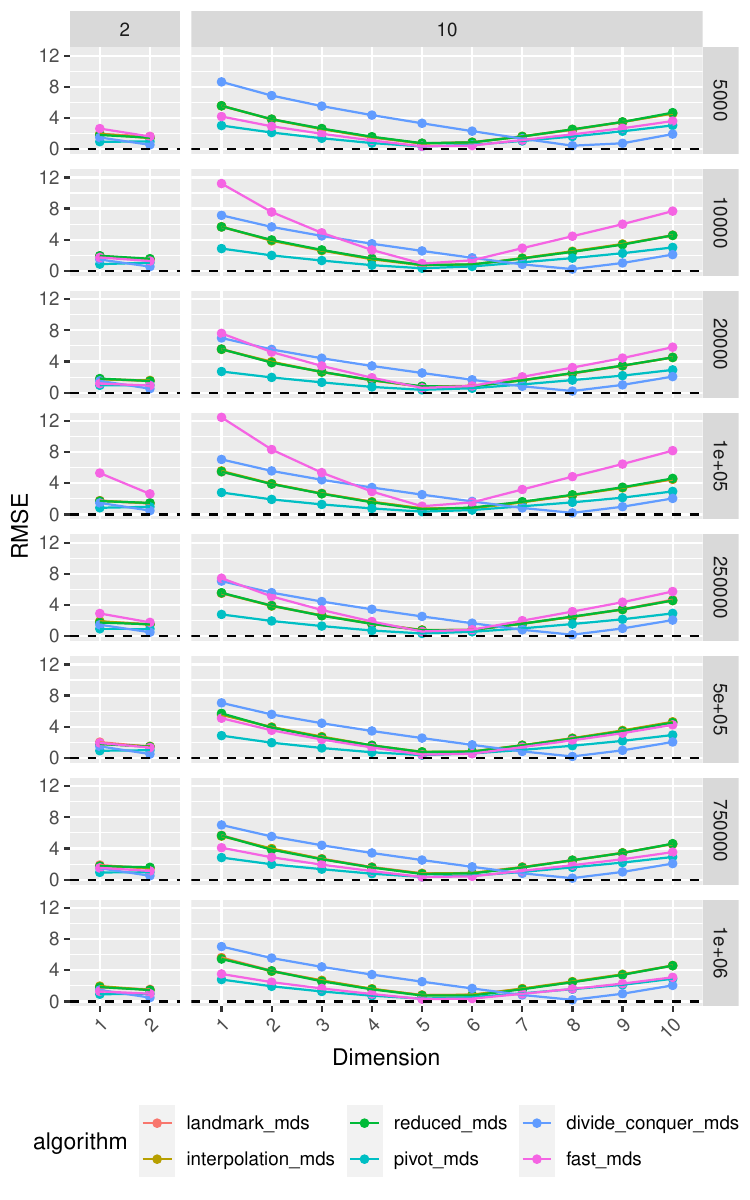}
	\caption{
		RMSE of the estimators for the variance of the dominant dimensions.
	}
	\label{fig:rmse}
\end{figure}

In this section we study how the eigenvalues provided by the algorithms estimate the variance of the dominant directions. Since these variances were equal to 15, it was expected the eigenvalues to be close to 15. 
Figures \ref{fig:bias} and \ref{fig:rmse} display the bias and the RMSE, respectively, taking into account the number of dominant dimensions, the dominant dimension and the sample size.

\blue{Pivot MDS had the lowest bias and RMSE, followed by LMDS, interpolation MDS, RMDS (these three being almost indistinguishable), then fast MDS and, finally, divide-and-conquer MDS, which systematically overestimated. Observe that fast MDS had a strange behavior for sample size $100 000$.}

In Figure \ref{fig:bias} it can be seen that the bias was positive for the first dominant dimensions and negative for the last ones. 
This happened because, when performing MDS with a particular sample, the estimated eigenvalues were sorted in decreasing order. 
Then, given that the theoretical value of \blue{all the eigenvalues} were the same, 
the first estimated ones tended to be larger than the true value, and the last ones tended to be smaller. This bias trend had effects on RMSE, as shown in Figure \ref{fig:rmse} where it can be seen that the RMSE was lower for intermediate dimensions, those having bias close to zero.



\subsection{Time to obtain an MDS configuration}\label{sect:simul.time}
\begin{figure}
	\centering
	\includegraphics[width=\textwidth]{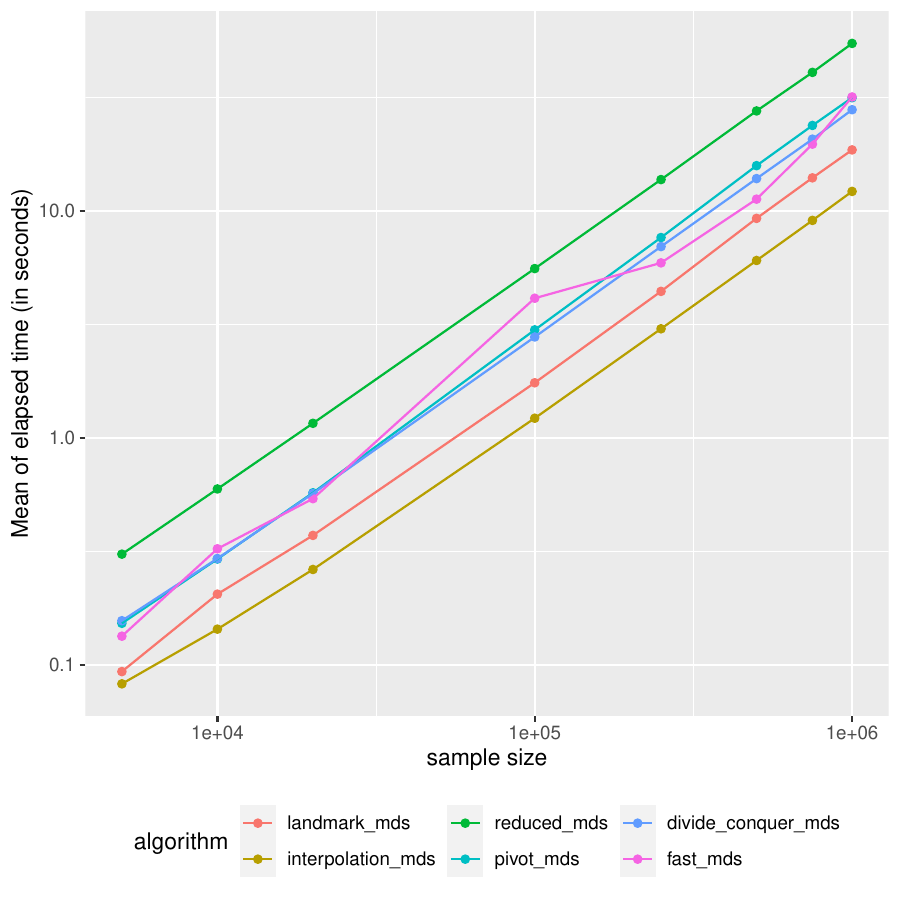}
	\caption{
Log-log plot of the mean of elapsed time against sample size. Colors represent the MDS method.
	}
	\label{fig:time_mds}
\end{figure}

In this section we study the cost of each algorithm in terms of speed.
Figure \ref{fig:time_mds} represents the log-log plot of mean time (in seconds) needed to obtain an MDS configuration as a function of the sample size (horizontal axis) and the MDS method (color). 
\blue{For all the sample sizes, the fastest algorithm was interpolation MDS, followed by LMDS, divide-and-conquer MDS, fast MDS, pivot MDS and finally RMDS. 
Note that divide-and-conquer MDS, fast MDS and pivot MDS behaved similarly in terms of elapsed times.
Additionally, it can be observed that execution time was approximate linear in sample size for the six algorithms. 
}

As a particular case, Table \ref{table:time_1000000_100_10} contains the $2.5\%$ quantile ($\q_{0.025}$), the mean value ($\mean$) and the $97.5\%$ quantile ($\q_{0.975}$)  for the elapsed time related to the scenario which sample size was $1 000 000$, there were 100 columns ($k = 100$) and 10 dominant dimension ($h=10$). The results related to the quantiles for the remaining scenarios (graphics not included here) were similar to the ones in Table  \ref{table:time_1000000_100_10}.

\begin{table}
	\caption{
		Quantiles of order $2.5\%$ $(\q_{0.025})$ and  $97.5\%$ $(\q_{0.975})$, and $\mean$ 
		values for the elapsed time (in seconds) for the scenario which sample 
		size was $1 000 000$, there were 100 columns and 10 dominant dimension.
	}
	\label{table:time_1000000_100_10}
	\centering
	\blue{
	\begin{tabular}{lrrr}
		\hline
		Algorithm & $\q_{0.025}$ & $\mean$ & $\q_{0.975}$\\ 
		\hline
LMDS                   & 23.46 & 24.27 & 24.82\\
Interpolation MDS      & 18.21 & 18.34 & 18.48\\
RMDS                   & 91.74 & 92.20 & 93.01 \\
Pivot MDS              & 36.19 & 37.38 & 38.04 \\
Divide-and-conquer MDS & 44.57 & 45.06 & 45.74 \\
Fast MDS               & 61.51 & 61.74 & 61.97 \\
		\hline
	\end{tabular}}
\end{table}

%

\section{Using MDS algorithms with EMNIST data set}
\label{sect:EMNIST}
In this section we {have used} \blue{all the} MDS algorithms with a real large data set: EMNIST (\citeNP{emnist}) available at
\url{https://www.nist.gov/itl/products-and-services/emnist-dataset}. 
The EMNIST data set is composed by gray-scaled handwritten character digits,  lowercase letters  and capital letters. They are derived from the Special Database 19 (\citeNP{nist_special})  and converted to a $28 \times 28$ pixel image format. The images have this size so that they match with the  MNIST data set format (\citeNP{mnist}).  In total, there are $814255$ images divided into  62 classes: 
10 digits (from `0' to `9'; the 49.5\% of the total), 
26 lowercase letters (from `a' to `z'; 23.5\%) and 
26 capital letters (from `A' to `Z'; 27\%). 

The Euclidean distance between the vector representation of the images in dimension $28^2=784$ {was used} to perform MDS with the \blue{six} algorithms. \blue{We first computed the MDS configuration requiring a low dimension equal to 10 ($r = 10$). Then, we used the first 2 dimensions to visualize the result.  With regards  to $\ell$, it} was set in same way as described in Section \ref{sect:l_election}. Divide-and-conquer MDS and fast MDS required to specify two extra parameters: $c$ and $s$ respectively. They both were set to \blue{50}. 
In particular, the principal coordinates provided by the six algorithms are highly correlated with those 
eventually obtained from classical MDS (correlations above $0.98$). The most correlated results are those of LMDS, interpolation MDS, RMDS (which are almost indistinguishable), followed by divide-and-conquer MDS, pivot MDS and then fast MDS.
Table \ref{table:emnist_time_gof} displays the time needed to obtain an MDS configuration. \blue{Pivot MDS was the fastest algorithm closely followed by LMDS and interpolation MDS. Then, divide-and-conquer MDS and fast MDS were slower than the other three previous methods. Finally, RMDS took around 10 minutes to obtain an MDS configuration. }

\begin{table}
	\caption{
\blue{
		Time  (in seconds) required to obtained the low dimensional configuration.
	}}
	\label{table:emnist_time_gof}
	\centering
\blue{
	\begin{tabular}{lr}
		\hline
		Algorithm &Time \\ 
		\hline
LMDS                   					    & 96.87 \\
Interpolation MDS     	 		 &  98.55\\
RMDS									    &  618.67\\
Pivot MDS								 &  91.76\\
Divide-and-conquer MDS	  &  168.33\\
Fast MDS               					 &  205.24\\
		\hline
	\end{tabular}}
\end{table}

\begin{figure}
	\centering
	\includegraphics[width=\textwidth]{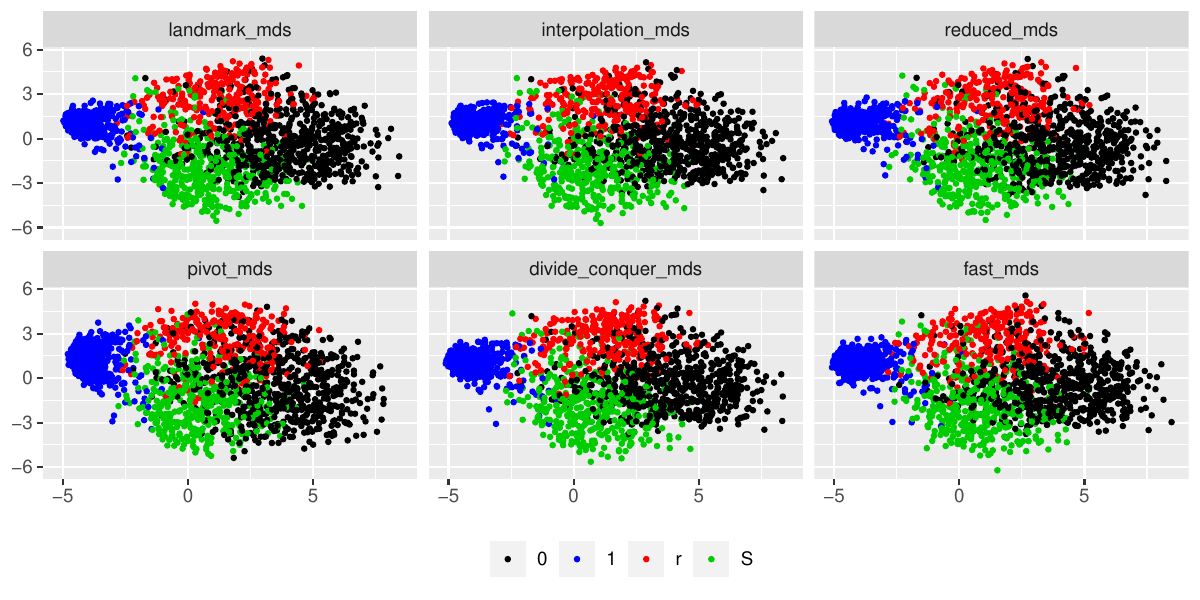}
	\caption{MDS configuration using all the algorithms.}
	\label{fig:emnist}
\end{figure}

Figure \ref{fig:emnist} shows the MDS configuration for each of the algorithms. In order to provide a comprehensive figure, \blue{we took a random sample of 2000 images from the following categories: `0', `1', `r' and `S'.} It can be seen that the \blue{six} methods gave similar results. \blue{The first dimension separated  `0' (rightmost part) from  `1' (leftmost part) and the second one separated 'r' (top part) from `S' (bottom part).}

\section{Conclusions}
\label{sect:concl}
\blue{In this work, we present six algorithms to obtain an MDS configuration for large data sets: 
LMDS, interpolation MDS, RMDS, pivot MDS, divide-and-conquer MDS and fast MDS.
Two of them (interpolation MDS and divide-and-conquer MDS)
are new proposals. 
We have proved that the distance-based triangulation used in LMDS coincides with the Gower interpolation formula used in interpolation MDS and RMDS.
In addition, we provide an \textsf{R} package that implements the \blue{six} MDS algorithms. 
}
	
\blue{We have developed an extensive simulation study to compare the performance of the six algorithms, both in terms of computational efficiency and the ability to recover the underlying low dimensional structure of the data.
According to the simulations, all the algorithms provide configurations similar to those eventually given by the classical MDS algorithm. 
In particular, the principal coordinates provided by the six algorithms are highly correlated with those 
hypothetically obtained from classical MDS (correlations above $0.98$). The most correlated results are those of LMDS, interpolation MDS, RMDS (which are almost indistinguishable), followed by divide-and-conquer MDS, pivot MDS and then fast MDS.
}

\blue{
Additionally, the six MDS algorithms provide good estimations of the principal coordinates variances, which are known by the simulation design, being pivot MDS the most accurate, followed by fast MDS, LMDS, interpolation MDS, RMDS and, finally, divide-and-conquer MDS. 
The performance of LMDS, interpolation MDS and RMDS is, again, practically coincident. 
}
	
\blue{Regarding the execution time, a clear difference between algorithms is observed in our simulation experiments.
Interpolation MDS is the fastest method, followed by LMDS, and RMDS is the slowest. Pivot MDS, divide-and-conquer MDS and fast MDS show similar execution time.}

\blue{As a final challenge for the algorithms, we have used them to obtain an MDS configuration for the real large data set EMNIST (more than $800000$ points). 
Since classical MDS algorithm {could} not be used with this data set, we {did} not have a gold standard to compare against.
The time needed to obtain the low dimensional configurations is admissible 
(always below 3.5 minutes, except for the RMDS, which takes $10.3$ minutes).
In this example, pivot MDS, LMDS and interpolation MDS are the fastest methods, with elapsed times around $1.5$ minutes.}

\blue{As a global conclusion, the \blue{six} algorithms are suitable for obtaining low dimensional configurations for large data sets, but we recommend to use interpolation MDS, 
for several reasons: 
(i) it is the fastest method in simulations,
(ii) it provides satisfactory results (almost identical to those of LMDS and RMDS), and
(iii) it does not incur in memory problems (an issue that could arise in LMDS and pivot MDS because both require storing a distance matrix of size $\ell\times n$).}

\backmatter

\section*{Acknowledgments}
This research was supported by the Spanish Research Agency (AEI)
under project PID2020-116294GB-I00, 
by AGAUR under grants 2020 FI SDUR 306 and 2021 SGR 00613,
and by UPC under AGRUPS-2022.

\bibliography{bibliography}


\end{document}